# The Case Against Hall-Petch Hardening in High Entropy Carbide Ceramics

Ali Sarikhani[a,*], Ana C. Feltrin[a], Gregory E. Hilmas[b], David W. Lipke[b], Douglas E. Wolfe[c], Stefano Curtarolo[d,†], Shen J. Dillon[e], William G. Fahrenholtz[b,*]

[a] *Missouri University of Science and Technology, Materials Research Center, Rolla, MO, USA*
[b] *Missouri University of Science and Technology, Materials Science and Engineering, Rolla, MO, USA*
[c] *Pennsylvania State University, Materials Science and Engineering, University Park, PA, USA*
[d] *Duke University, Mechanical Engineering and Materials Science, Center for Extreme Materials, Durham, NC, USA*
[e] *University of California, Irvine, Materials Science and Engineering, Irvine, CA, USA*

***ORCID:***

***Sarikhani**: 0000-0001-8121-0867*
***Feltrin**: 0000-0001-9920-1643*
***Hilmas**: 0000-0002-1611-5457*
***Lipke**: 0000-0002-4557-6690*
***Wolfe**: 0009-0004-5090-0290*
***Curtarolo**: 0000-0003-0570-8238*
***Dillon**: 0000-0002-6192-4026*
***Fahrenholtz**: 0000-0002-8497-0092*

## Abstract

Grain size is often used with the Hall-Petch relationship to justify differences in hardness in ceramic materials. Herein, hardness does not vary systematically with grain-size for fully dense, single-phase (Cr,Mo,Ta,V,W)$C_{1-\delta}$ high entropy carbide ceramics for grain sizes that varied by a factor of three. Fully dense, single-phase ceramics with grain sizes from 9.3±0.3 to 28.8±0.7 µm exhibited a pronounced indentation size effect, with Vickers hardness decreasing from ~28-30 GPa at 0.49 N to ~20-21 GPa at 9.81 N, and Berkovich nanohardness ranging from 26 to 30 GPa at 10 mN. However, at a given load, hardness remained within a narrow range across the grain-size series, and no consistent Hall-Petch dependence was resolved. The lack of grain-size dependence likely indicates that the deformation volume sampled by the indenter was not controlled by grain-boundary interactions; instead, hardness was governed primarily by indentation load and local response of the rock salt carbide matrix.

* Corresponding authors: A. Sarikhani (as5kw@mst.edu), W. G. Fahrenholtz (billf@mst.edu)

† Stefano Curtarolo was an Editor of the journal during the review period of the article. To avoid a conflict of interest, Stefano Curtarolo was blinded to the record and another editor processed this manuscript.

High entropy carbide (HEC) ceramics have the potential to combine melting temperatures above 3000°C with the ability to tailor properties using flexible multi-principal element design [1-6]. Previous reports have described favorable properties such as strength retention to higher temperatures [7] and enhanced oxidation resistance [8-10] compared to carbides of individual metals. Several previous studies have focused specifically on the hardness of HECs [1,11-14]. Reported hardness values for high entropy carbides vary widely, with prior studies showing nanoindentation or Vickers hardness values from approximately 19 GPa to more than 40 GPa depending on composition, indentation load, valence electron concentration, lattice strain, and solid-solution effects [1,11-13]. For the $(Cr,Mo,Ta,V,W)C_{1-\delta}$ system, our previous work showed that Vickers hardness at 0.49 N remained nearly constant at approximately 27-28 GPa despite variations in carbon content and grain size, indicating that hardness was not strongly controlled by these processing-related changes in that study [14].

Grain size is a relevant microstructural parameter for mechanical behavior. Hardness in ceramics is often discussed in terms of grain-size-dependent strengthening, commonly through Hall-Petch-type relations with a linear relationship between yield strength and inverse square root of grain size [15-19]. Load-dependent effects (i.e., the indentation size effect) are well known in ceramics, and their fundamental origins have been discussed in prior mechanistic studies [16,20]. Some prior studies have observed grain size effects on hardness at constant load [15,16], but recent studies on HECs have attributed differences in hardness to the Hall-Petch effect without providing sufficient analysis to determine if the differences were due to a measurable grain-size effect, or from concurrent changes in density, phase constitution, defect state, or indentation load [11-14].

For previous studies of HEC ceramics, isolating the effect of grain size on hardness has remained challenging. Pötschke et al. prepared dense rock salt HECs by sinter-HIP, vacuum sintering, and SPS and reported that the finer-grained (Hf,Ta,Nb,Ti,V)C specimens generally showed higher hardness than the coarser-grained (Ta,Nb,Ti,V,W)C specimens; however, within the finer-grained composition, the smallest-grained sinter-HIP specimen did not have the highest hardness, indicating that hardness was not controlled by grain size alone [21]. Petrus et al. further showed that dense single-phase (Hf,Ta,Zr,Nb,Ti)C and (Mo,Nb,Ta,V,W)C exhibited a strong indentation size effect, with load-independent hardness values depending on the load range used for the analysis [22]. More recently, Hu et al. showed that two-step sintering of (Nb,V,Ta,Mo,W)C improved density and retained a smaller grain size, while increased sintering temperature and longer holding time promoted grain growth and weakened mechanical performance [23]. These studies show that grain size, density, composition, sintering history, and indentation load can all affect reported hardness values, motivating the present isochemical study of a dense, single-phase HEC series. Accordingly, one specific need is to evaluate grain-size-dependent hardness in a fully dense, single-phase HEC system under conditions where microstructure can be varied without major changes in phase constitution or porosity. Spark plasma sintering (SPS) is a useful route to address this issue because a controlled temperature series can generate systematic grain-size variations while maintaining dense, nominally single-phase ceramics [21,24]. (Cr,Mo,Ta,V,W)C was selected as a model HEC because of its single-phase rock-salt formability, high valence-electron concentration, and associated combination of metallic bonding character, fracture resistance, and high hardness [25-28]. Its thermophysical properties and grain-growth behavior have also been established in our previous studies [14,29]. The present work challenges conventional thinking in the field by showing that a distinct Hall-Petch dependence was not resolved in the indentation hardness of fully dense (Cr,Mo,Ta,V,W)$C_{1-\delta}$ ceramics over the controlled grain-size range investigated.

The starting powders for the (Cr,Mo,Ta,V,W)$C_{1-\delta}$ ceramics were $Cr_2O_3$ (99.5%, 0.7 µm; Elementis), $MoO_3$ (99.9%, 6 µm; US Research Nanomaterials), $Ta_2O_5$ (99.8%, 1-5 µm; Atlantic Equipment Engineers), $V_2O_5$ (99.6%, -10 mesh; Alfa Aesar), $WO_3$ (99.9%, ~80 nm; Inframat Advanced Materials), and carbon black (C120; Cabot). The starting materials were batched to have an equimolar metal content as described in our previous research according to the 7.5 wt% carbon deficient composition [14]. The term 7.5 wt.% carbon deficient refers to the carbon content in the starting powder batch compared to the calculated stoichiometry of the carbothermal reduction reaction and is not meant to imply any specific carbon vacancy content in the final ceramic. This composition previously produced dense, nominally single-phase rock-salt ceramics without obvious residual oxides or free carbon [14]. Powders were blended by high-energy ball milling in a WC vial using WC media with a media-to-powder mass ratio of 15:2, then sieved through 60 mesh and compacted into pellets.

The pellets underwent carbothermal reduction at 1610 °C for 3 h under a mild vacuum (~13.3 Pa) in a graphite-element furnace (HP50-7010 G, Thermal Technology). The reacted material was crushed, passed through a 100-mesh sieve, divided into five portions, and densified by spark plasma sintering (SPS; DCS10, Thermal Technology, LLC., USA) at temperatures between 1750 and 1950 °C. The SPS cycle began under vacuum (<6 Pa) with heating to 1600 °C at 100 °C/min under a 15 MPa applied pressure, followed by a 5 min dwell. The uniaxial pressure was then increased to 50 MPa, and the temperature was raised to the final densification temperature at 100 °C/min. The densification temperatures ranged from 1750°C to 1950°C to produce different final grain sizes while reaching full density. After a 10 min dwell at the respective peak temperatures, the pressure was reduced to 25 MPa during cooling to 1200 °C at 50 °C/min.

After SPS, portions of each sintered specimen were crushed into powder for phase analysis by room-temperature X-ray diffraction (XRD; X'Pert MPD, Philips). The powders were mixed with an $\alpha$-$Al_2O_3$ internal standard (99.8%, D50 ~0.6 µm; Almatis) and analyzed by Rietveld refinement to determine the lattice parameters. Bulk density was measured by the Archimedes method, and theoretical density was calculated assuming a rock salt carbide structure using the refined lattice parameter together with the average metal-site composition obtained from energy-dispersive spectroscopy (EDS). Relative density was calculated from the ratio of bulk to theoretical density, and the uncertainty was estimated using error propagation of bulk and theoretical densities. Uncertainty in the EDS quantification was the dominant contributor to the propagated relative density error. Based on the measured mass loss of the WC milling media before and after milling, each sample incorporated an average of ~0.03 g WC contamination; its effect on the theoretical density was calculated to obtain corrected theoretical density and relative density values.

Micrographs used for grain-size analysis were obtained by scanning electron microscopy (SEM; Axia ChemiSEM, Thermo Fisher Scientific Inc.). Grain boundaries were manually traced using image manipulation software (GIMP, The GNU Image Manipulation Program Team [30]), and Feret diameters were measured using image analysis software (ImageJ; National Institutes of Health, Bethesda, MD). For each sintering condition, 200-500 grains were analyzed and the uncertainty was reported as the standard error of the mean. Additional SEM imaging and EDS analysis, including elemental mapping and point quantification, were performed using a second SEM system (PIONEER, Raith 150 eLine Plus).

Vickers hardness was measured on polished cross-sections using a Duramin 5 microhardness tester (Struers, Cleveland, OH) in accordance with ASTM C1327. Indentations were made using loads from 0.49 to 9.81 N with a dwell time of 15 s. For each load, at least five valid indents were measured using an optical microscope (KH-3000, Hirox-USA), and hardness values were calculated from the average diagonal lengths of the impressions. Berkovich nanohardness (Hysitron TI 980 TriboIndenter, Bruker, Germany) and the reduced modulus were obtained by a load controlled 6 × 6 basic quasistatic trapezoid array of indents, with an 8 μm spacing, 10 mN indent load with a 5-2-5 second load-hold-unload configuration on polished cross-sections. Nanoindentation measurements exhibiting abnormal load-displacement behavior associated with pores, defects, or surface features were excluded from the analysis. In addition, the three highest and three lowest values were excluded to minimize the influence of anomalous measurements. Reported uncertainties for Vickers hardness, nanohardness, and reduced modulus refer to standard deviations.

Dense, single-phase (Cr,Mo,Ta,V,W)$C_{1-\delta}$ ceramics with systematically different grain sizes were produced by varying the SPS temperature (Table 1). The specimen designations, sintering temperatures, lattice parameters, grain sizes, and relative densities for the five sintering conditions are summarized in Table 1:

| Sample | Sintering temperature (°C) | Lattice parameter (Å) | Avg. grain size (μm) | Relative density (%) |
|---|---|---|---|---|
| **S1750** | 1750 | 4.2767 ± 0.00015 | 9.3 ± 0.3 | 100 ± 3 |
| **S1800** | 1800 | 4.2773 ± 0.00009 | 13.3 ± 0.5 | 100 ± 2 |

| S1850 | 1850 | 4.2775 ± 0.00001 | 18.7 ± 0.5 | 102 ± 3 |
|---|---|---|---|---|
| **S1900** | 1900 | 4.2793 ± 0.00010 | 21.5 ± 0.7 | 100 ± 3 |
| **S1950** | 1950 | 4.2806 ± 0.00003 | 28.8 ± 0.7 | 98 ± 2 |

**Table 1.** Specimen designations, sintering temperatures, lattice parameters, average grain sizes, and relative densities for the fully dense (Cr,Mo,Ta,V,W)$C_{1-\delta}$ ceramics used for hardness evaluation.

The lattice parameter increased with SPS temperature from 4.2767 Å for S1750 to 4.2806 Å for S1950. The increase may be due to changes in chemical homogeneity, local strain, and/or C-site sublattice occupancy. Because the final C and O contents were not independently quantified, the expansion cannot be assigned to a specific defect mechanism.

Grain size increased from 9.3 ± 0.3 µm for S1750 to 28.8 ± 0.7 µm for S1950, providing a well-defined microstructural series for evaluating hardness. Room-temperature XRD analysis indicated that all ceramics retained the single-phase rock salt structure; an enlarged view of the (111) reflection showing the systematic peak shift with sintering temperature is provided in Figure S1 of the Supplementary Material, while the complete diffraction patterns were reported previously [29]. Representative SEM micrographs in Figure 1 show progressive grain coarsening with increasing sintering temperature. A representative SEM micrograph and EDS elemental maps for S1750, illustrating broadly homogeneous elemental distributions, are provided in Figure S2 of the Supplementary Material; EDS maps for all the specimens were reported previously [29]. No obvious residual porosity was observed by SEM.

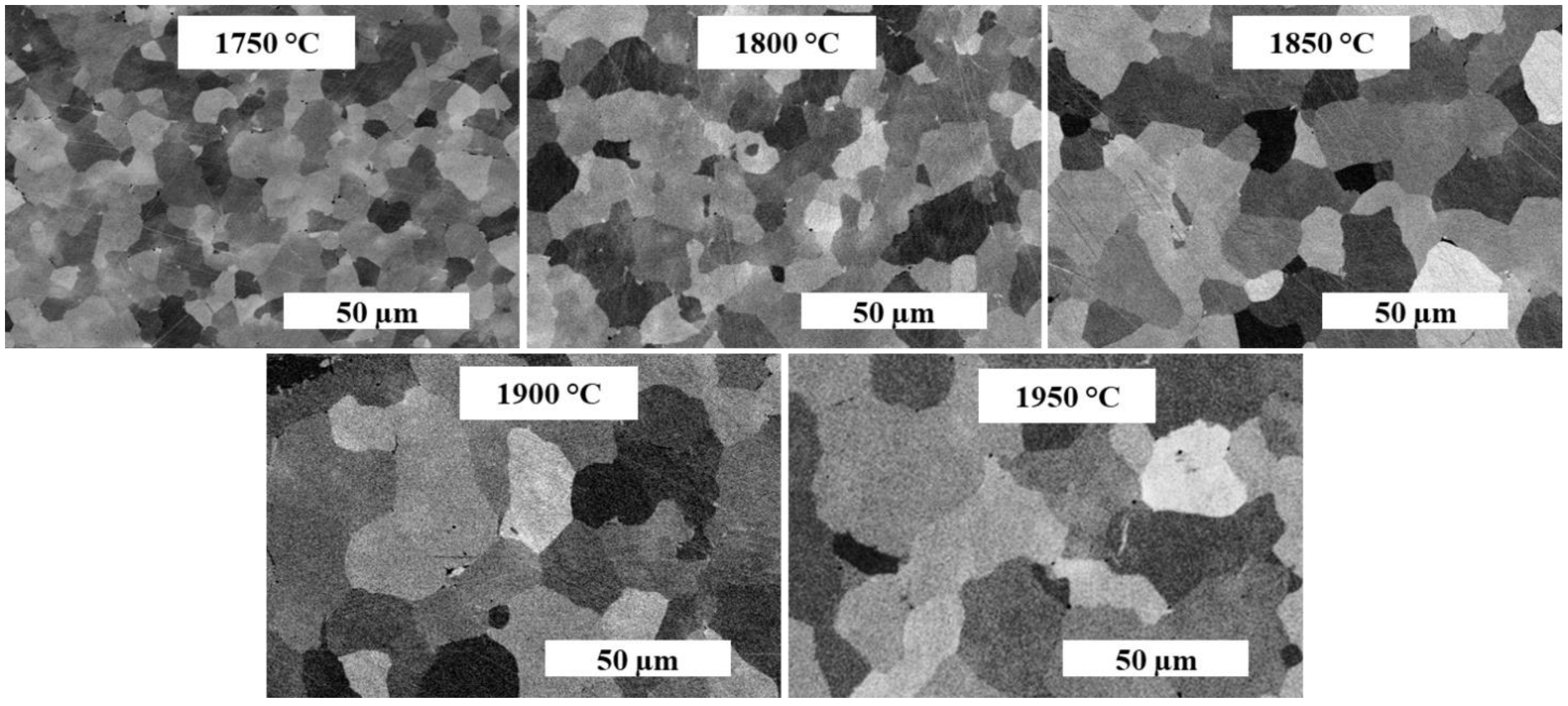

**Figure 1.** Representative SEM micrographs for (Cr,Mo,Ta,V,W)$C_{1-\delta}$ ceramics sintered at 1750, 1800, 1850, 1900, and 1950 °C. The SEM images show progressive grain coarsening with increasing sintering temperature.

When the Vickers hardness (Figure 2) and nanohardness (Figure 3) data were plotted as a function of grain size at fixed loads, hardness appeared to be independent of grain size over the range that was produced. Grain size increased by more than a factor of 3 from 9.3 to 28.8 μm, yet the hardness values at each load remained within a relatively narrow band. For completeness, the full hardness versus load behavior for all specimens, which showed the pronounced indentation size effect, is provided in the supplementary material (Figure S3). At 0.49 N, the measured Vickers hardness values were all near ~28-30 GPa, while at 9.81 N they were all near ~20-21 GPa. Over the same range of loads, the average indent diagonal increased from approximately 5.5-5.7 μm at 0.49 N to approximately 29.7-30.3 μm at 9.81 N, corresponding to indent-size/grain-size ratios of approximately 0.19-0.62 at 0.49 N and 1.05-3.24 at 9.81 N (Figure S4). Intermediate loads showed similarly limited variations with grain size. For comparison, the reduced modulus obtained from the same Berkovich nanoindentation measurements varied from ~276 to 328 GPa with no correlation to grain size. The modulus changes were similar to those of nanohardness, suggesting that the differences were due local structural or microstructural variations rather than a systematic grain-size effect, as shown in Figure S6. Crystallographic orientation could also potentially contribute to hardness variations, particularly in nanoindentation in which the indents are typically smaller than the grain size of the ceramics. In the present case, the orientation dependence of hardness would be weak due to the crystallographic symmetry of the cubic rock salt structure of the carbides compared to lower symmetry materials where a pronounced effect has been reported [31-33]. In addition, the reported values from nanoindentation in the present study were the average of an array of 36 indents, which presumably represent a statistical sampling of different orientations for each material.

Nanoindentation was performed using a Berkovich diamond tip at a maximum load of 10 mN. The characteristic lateral contact dimension was approximately 0.9 µm, with average maximum indentation depths of approximately 138-158 nm, substantially smaller than the investigated average grain sizes. Representative Vickers impressions at 0.49 and 9.81 N and a representative Berkovich nanoindentation array are shown in Figure S5; no obvious gross chipping or extensive surface damage was observed, although radial corner cracking occurred at the highest Vickers load. At 9.81 N, the indent dimensions were comparable to or larger than the average grain size for all specimens. Thus, although the microstructure coarsened systematically with increasing densification temperature, neither the Vickers nor nanohardness showed a consistent trend with $G^{-1/2}$, where G being the grain size, indicating that a distinct Hall-Petch contribution to hardness could not be resolved within the investigated grain size range.

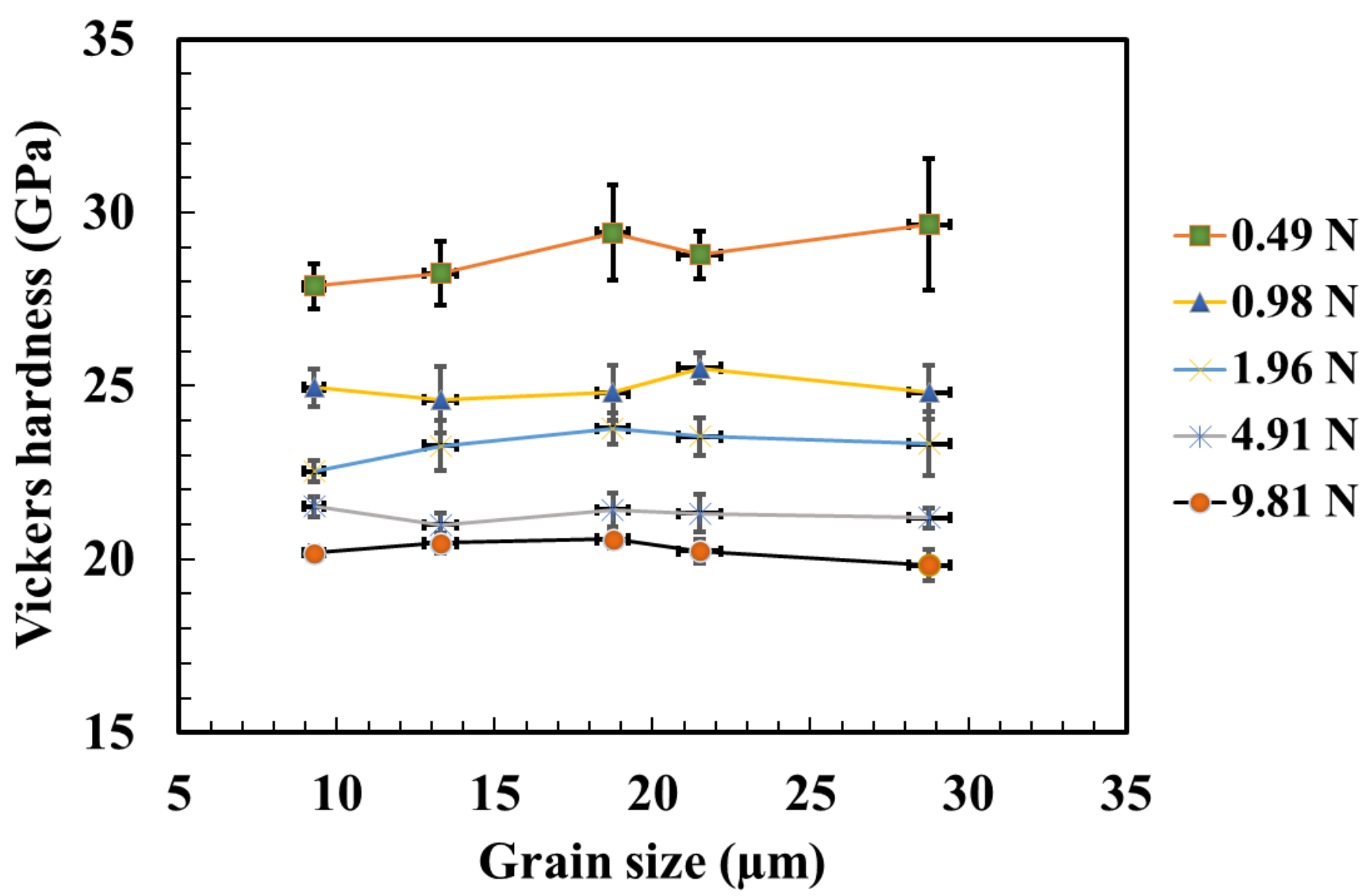

**Figure 2.** Hardness as a function of grain size for fully dense (Cr,Mo,Ta,V,W)$C_{1-\delta}$ ceramics at indentation loads of 0.49, 0.98, 1.96, 4.91, and 9.81 N.

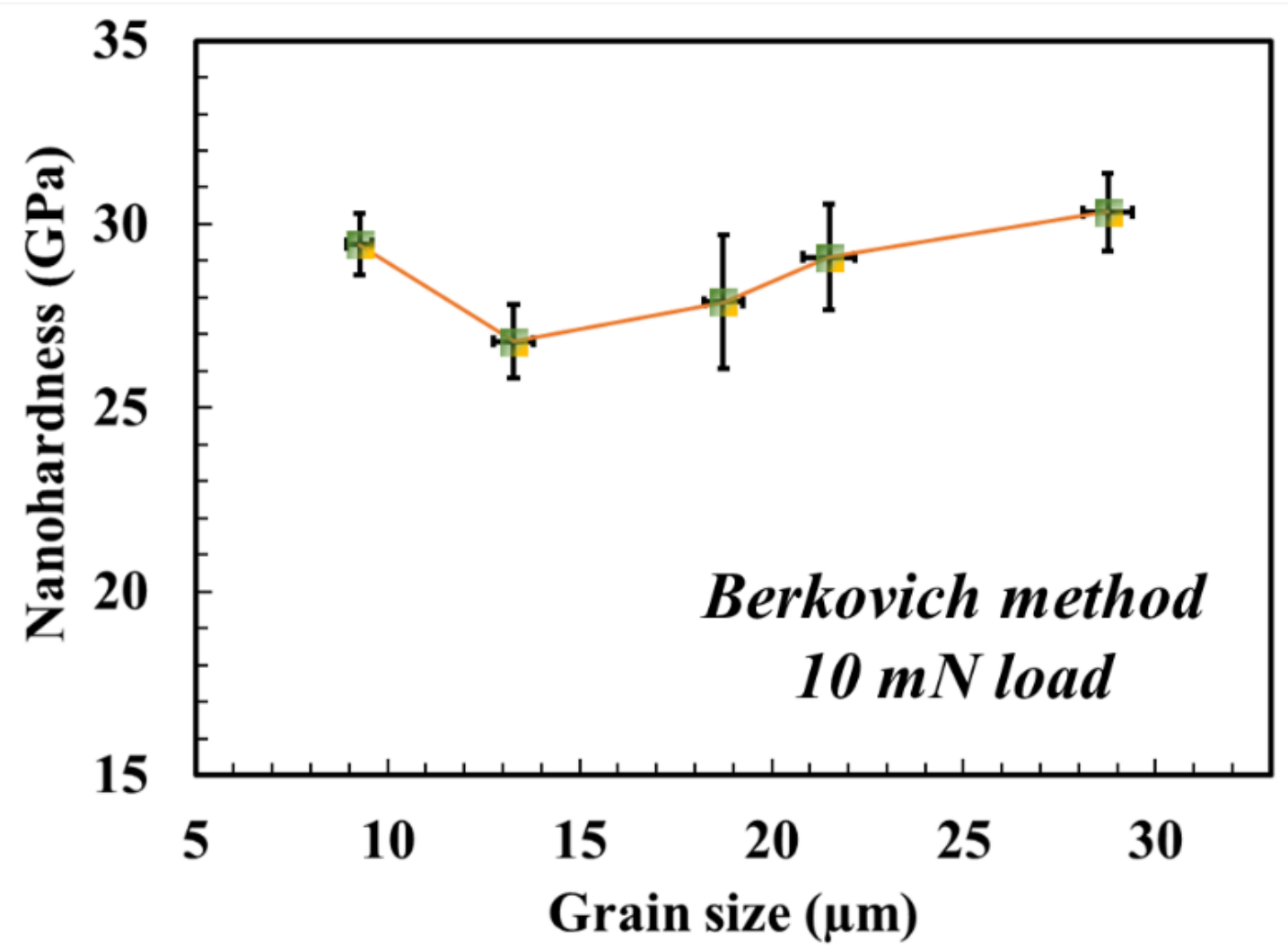

**Figure 3.** Nanohardness as a function of grain size for fully dense (Cr,Mo,Ta,V,W)$C_{1-\delta}$ ceramics measured using a Berkovich indenter at a 10 mN load.

Although grain size is often reported to influence hardness of ceramics through Hall-Petch-type strengthening, the present results show that the grain size effect was not significant over the range of grain sizes that was investigated. One possible reason is that the grain-size window examined here, while sufficient to produce clear microstructural coarsening, may not have been large enough to generate a strong hardness difference relative to the measurement scatter and the pronounced indentation size effect. Quantitative comparisons with related rock salt carbides suggest that a Hall-Petch-type contribution over the present grain-size range would be difficult to resolve. For example, Ko et al. reported a Hall-Petch-type relation of $H=10.42+2.55G^{-1/2}$ for ZrC coating layers, where the hardness H is in GPa and the grain size G is in μm [34]. Applying this coefficient to the present grain-size range of 9.3-28.8 μm gives an estimated hardness change of only ~0.36 GPa. Similarly, Wang et al. reported that reducing the grain size of single-phase rock salt (Hf,Zr,Ta,Nb,Ti)C from 16.5 μm to the submicron range increased hardness only modestly at 9.8 N [35]. Based on the reported hardness values of 16.21 ± 1.04 and 17.07 ± 0.54 GPa, this corresponds to an apparent coefficient of ~0.8 $GPa \cdot \mu m^{1/2}$ and an estimated hardness change of only ~0.1 GPa over the grain-size range examined here. These comparisons indicate that any Hall-Petch-type contribution expected over the present grain-size interval would likely be much smaller than the measurement scatter and the pronounced indentation size effect. Hence, Hall-Petch strengthening or hardening is unlikely to dominate hardness differences in carbide ceramics with grain sizes on the order of microns to tens of microns.

Hall-Petch-type behavior in ceramics is often discussed by analogy to the original Hall-Petch relationship developed for metals [36]. Although the atomistic deformation mechanisms may differ between metals and ceramics, the key requirement in both cases is that grain-boundary interactions influence the deformation response, producing a dependence on $G^{-1/2}$ [19]. In the present work, however, the measured hardness did not scale with $G^{-1/2}$ over the investigated grain-size range, indicating that grain-boundary-controlled strengthening was not resolved under these indentation conditions. For comparison, conventional low-carbon steel with a Hall-Petch coefficient of ~600 $MPa \cdot \mu m^{1/2}$ predicts an increase in yield strength of approximately 40% when grain size decreased from 28.8 to 9.3 μm, from ~212 to ~297 MPa [37]. In contrast, no systematic hardness increase was resolved over the same grain-size range in the high-entropy carbide ceramics in the present study. For high entropy ceramics, once dense single-phase microstructures are achieved, hardness in this chemically complex carbide may be governed more strongly by intrinsic lattice resistance associated with multi-principal-element solid-solution effects and defect-related strengthening than by the change in grain-boundary area alone. However, the absence of a clear Hall-Petch trend in the present hardness data does not necessarily imply an absence of dislocation activity during indentation. Rather, it indicates that any grain-boundary strengthening contribution was not resolved over the investigated grain-size range. This may reflect the indentation length scale: at low loads, the deformation volume may have been largely confined within individual grains for many specimens, whereas at high loads the indent dimensions became comparable to or larger than the grain size, but the hardness response was dominated by the indentation size effect rather than by a systematic Hall-Petch-type trend. Overall, the results herein indicate that differences in hardness observed for carbide ceramics with different grain sizes are unlikely to be due to Hall-Petch strengthening or hardening when the grain sizes are on the order of microns or tens of microns.

**Acknowledgements**

The authors acknowledge Advanced Materials Characterization Core and the Materials Research Center of the Missouri University of Science and Technology for providing the facilities and resources used in this work. The authors thank Drs. Xiomara Campilongo, Simon Divilov, Hagen Eckert, Paolo De Angelis, and Scott Thiel from Duke University for their helpful discussion. This research was supported by the Office of Naval Research under Award No. N00014-24-1-2768.

**A. Sarikhani** (Conceptualization, Methodology, Visualization, Validation, Writing - original draft, Data curation), **A. C. Feltrin** (Methodology, Validation, Writing - review & editing, Data curation), **G. E. Hilmas** (Conceptualization, Supervision, Validation, Writing - review & editing), **D. W. Lipke** (Conceptualization, Validation, Writing - review & editing), **D. E. Wolfe** (Conceptualization, Writing - review & editing), **S. Curtarolo** (Conceptualization, Writing - review & editing), **S. J. Dillon** (Conceptualization, Writing - review & editing), **W. G. Fahrenholtz** (Conceptualization, Methodology, Supervision, Validation, Writing - review & editing).

The author Stefano Curtarolo is an Editorial Board Member for Acta and Scripta Materialia and was not involved in the editorial review or the decision to publish this article.

The data that support the findings of this study are available from the corresponding author upon request.

# *Supplementary Material*

## The Case Against Hall-Petch Hardening in High Entropy Carbide Ceramics

Ali Sarikhani[1,*], Ana C. Feltrin[1], Gregory E. Hilmas[2], David W. Lipke[2], Douglas E. Wolfe[3], Stefano Curtarolo[4,†], Shen J. Dillon[5], William G. Fahrenholtz[2,*]

[1] *Missouri University of Science and Technology, Materials Research Center, Rolla, MO, USA*
[2] *Missouri University of Science and Technology, Materials Science and Engineering, Rolla, MO, USA*
[3] *Pennsylvania State University, Materials Science and Engineering, University Park, PA, USA*
[4] *Duke University, Mechanical Engineering and Materials Science, Center for Extreme Materials, Durham, NC, USA*
[5] *University of California, Irvine, Materials Science and Engineering, Irvine, CA, USA*

***ORCID:***

***Sarikhani**: 0000-0001-8121-0867*
***Feltrin**: 0000-0001-9920-1643*
***Hilmas**: 0000-0002-1611-5457*
***Lipke**: 0000-0002-4557-6690*
***Wolfe**: 0009-0004-5090-0290*
***Curtarolo**: 0000-0003-0570-8238*
***Dillon**: 0000-0002-6192-4026*
***Fahrenholtz**: 0000-0002-8497-0092*

*S1. Crystal structure and lattice parameter evolution*

Room-temperature X-ray diffraction analysis indicated that all specimens retained the single-phase rock salt structure over the investigated sintering-temperature range. The complete diffraction patterns for these specimens were reported previously [29]. Figure S1 presents an enlarged view of the rock-salt (111) reflection near $2\theta \approx 36.4°$. With increasing SPS temperature, the reflection shifts systematically toward lower $2\theta$, consistent with the increase in refined lattice parameter from 4.2767 Å for S1750 to 4.2806 Å for S1950. Rietveld refinement was performed using $\alpha$-$Al_2O_3$ (99.8%, D50 ~0.6 µm; Almatis, Leetsdale, PA) as an internal standard. The small lattice-parameter increase may reflect changes in chemical homogeneity, local strain, and/or anion-sublattice occupancy. Because the final C and O contents were not independently quantified, the expansion cannot be assigned to a specific defect mechanism.

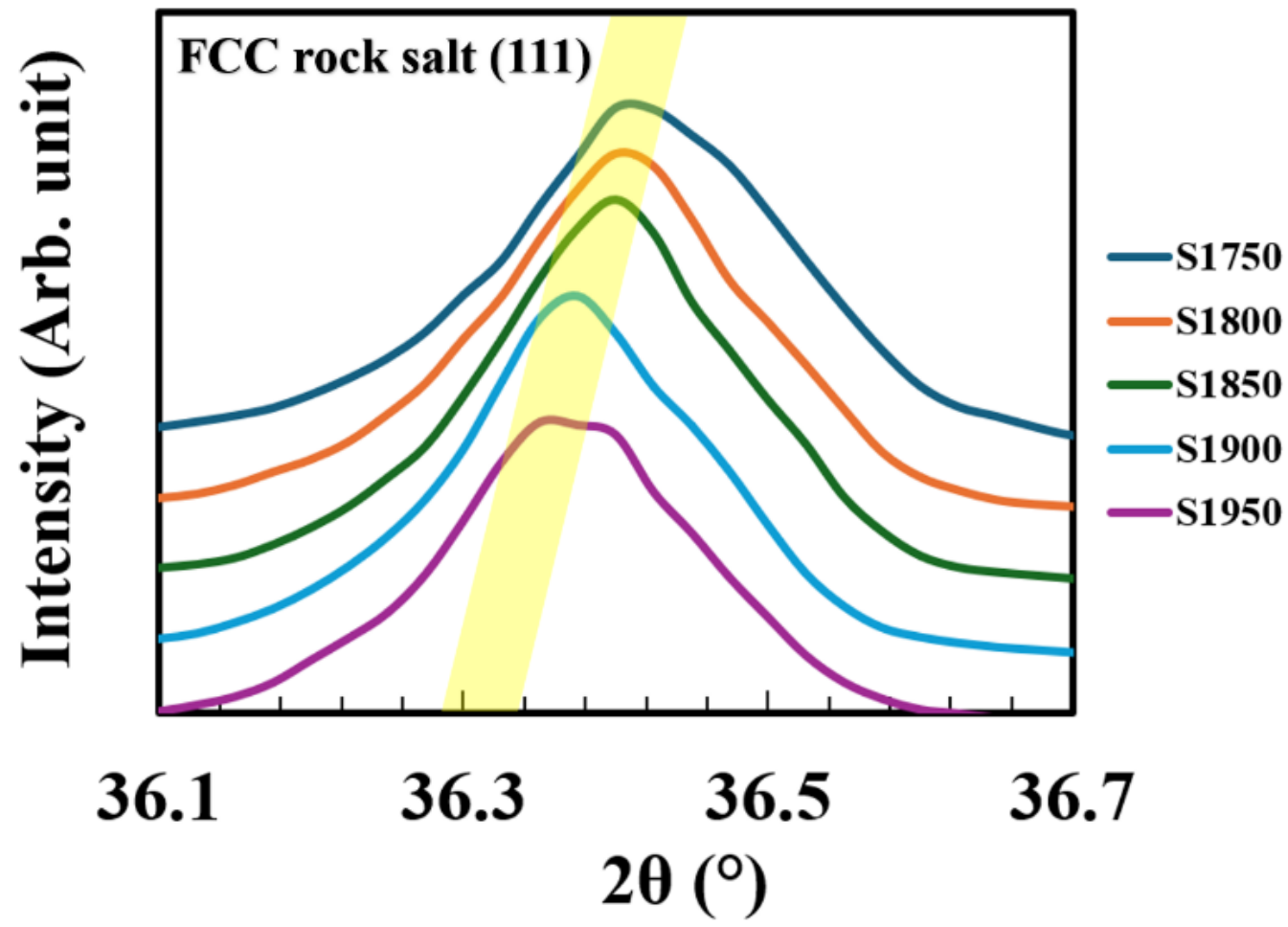

**Figure S1.** Enlarged view of the rock-salt (111) X-ray diffraction reflection of (Cr,Mo,Ta,V,W)$C_{1-\delta}$ ceramics sintered at 1750-1950 °C. The reflection shifts systematically toward lower 2θ with increasing SPS temperature, consistent with the measured increase in lattice parameter. The diffraction profiles are vertically offset for clarity.

*S2. Microstructure and composition*

Grain size increased from 9.3 ± 0.3 μm for S1750 to 28.8 ± 0.7 μm for S1950, providing a well-defined microstructural series for hardness evaluation. Progressive grain coarsening across the sintering-temperature series is shown in Figure 1 of the main text. Figure S2 presents a representative SEM micrograph and EDS elemental maps for S1750. The elemental maps show broadly homogeneous distributions of Cr, Mo, Ta, V, W, C, and O within the examined region. EDS maps for all the specimens were reported previously [29]. No obvious residual porosity was observed in the SEM observations. Although the calculated relative density of S1950 was 98 ± 2%, this value remained close to full density and no corresponding porosity was evident by SEM. Because grain size varied substantially whereas phase constitution and density remained essentially unchanged across the series, the hardness comparisons in the main text are interpreted in terms of microstructural length scale rather than effects related to porosity or second phases.

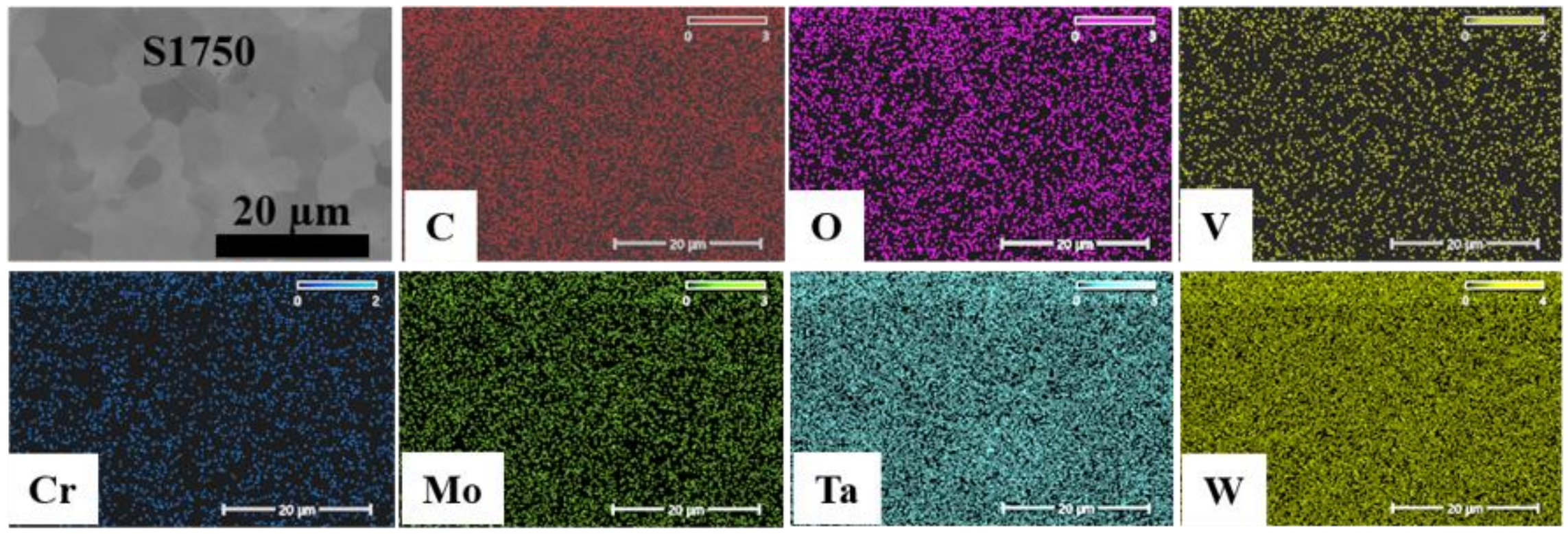

**Figure S2.** Representative SEM micrograph and EDS elemental maps for the S1750 (Cr,Mo,Ta,V,W)$C_{1-\delta}$ ceramic. The elemental maps show broadly homogeneous distributions of Cr, Mo, Ta, V, W, C, and O within the examined region.

*S3. Vickers hardness vs. load*

Hardness decreased with increasing indentation load for all specimens (Figure S3), consistent with the indentation size effect. At the lowest load (0.49 N), hardness values were in the range of ~28-30 GPa, whereas at the highest load (9.81 N) they converged to ~20-21 GPa. The overall load dependence was similar for all five grain sizes and the hardness values at a given load remained closely grouped. These results indicate that indentation load exerted a stronger influence on the measured hardness than the differences in grain size. The load dependence is consistent with prior ceramic hardness studies showing that hardness values measured at low loads are often higher than those measured at larger loads.

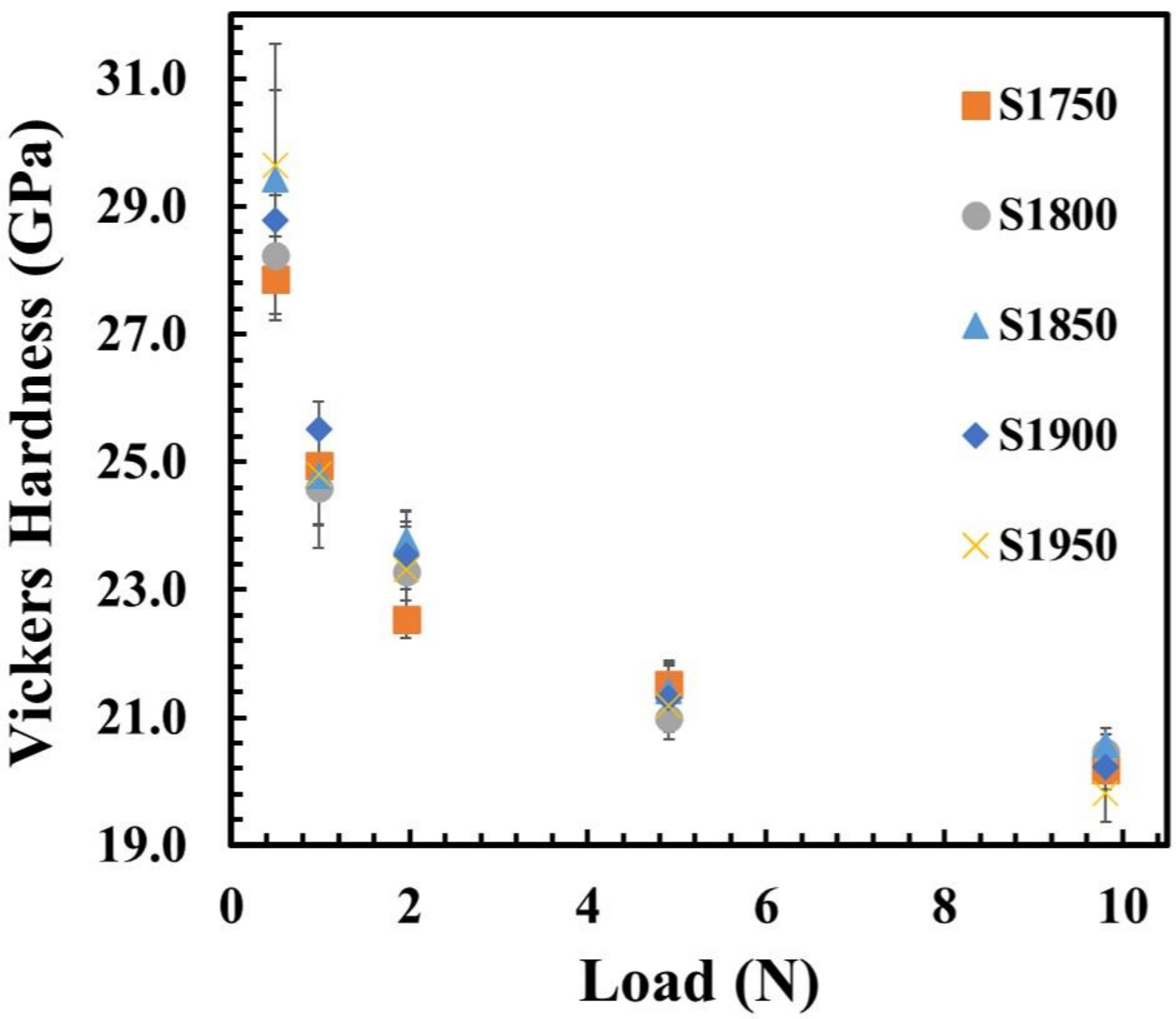

**Figure S3.** Hardness as a function of indentation load for fully dense (Cr,Mo,Ta,V,W)$C_{1-\delta}$ ceramics sintered at 1750-1950 °C.

*S4. Indentation size vs. grain size*

Figure S4 shows the average indent diagonal as a function of grain size for each Vickers indentation load. The average indent size was controlled primarily by the applied load and remained nearly constant across the grain-size series at a given load. The indent diagonal increased from approximately 5.5-5.7 μm at 0.49 N to approximately 29.7-30.3 μm at 9.81 N, corresponding to indent-size/grain-size ratios of approximately 0.19-0.62 and 1.05-3.24, respectively. Thus, the lowest-load indents were smaller than the average grain size for most specimens, whereas the highest-load indents were comparable to or larger than the average grain size. The error bars are small and may not be clearly visible in some cases because the measured indent diagonals were highly consistent for a given indentation load. Therefore, the lower-load measurements may have sampled deformation volumes that were too small to interact with multiple grain boundaries, while the higher-load measurements sampled regions comparable to or larger than the average grain size; the absence of a clear Hall-Petch-type trend should therefore be interpreted within this indentation-size/grain-size window.

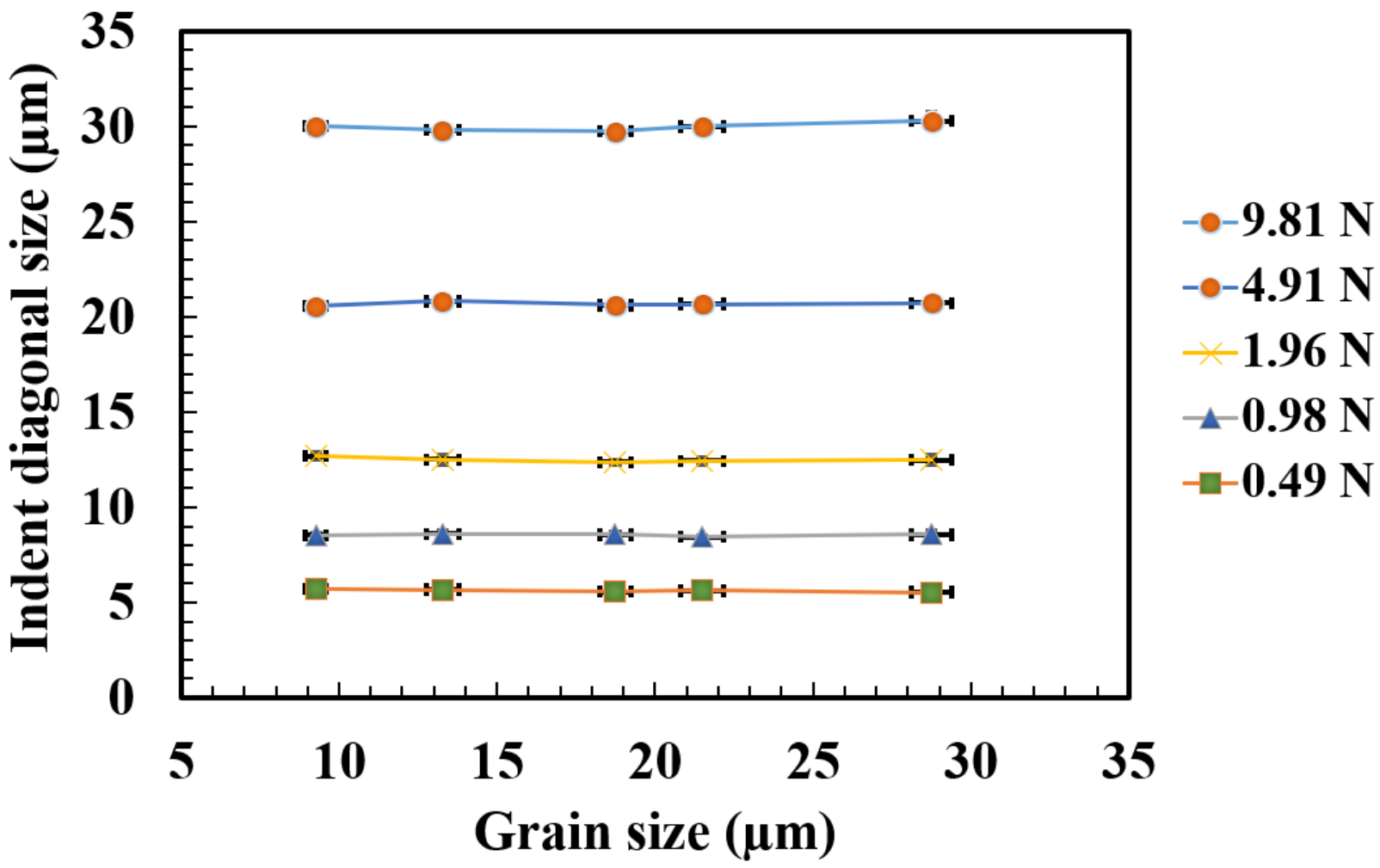

**Figure S4.** Average indent diagonal as a function of grain size for fully dense $(Cr,Mo,Ta,V,W)C_{1-\delta}$ ceramics at Vickers indentation loads of 0.49, 0.98, 1.96, 4.91, and 9.81 N. Error bars are small and may be obscured by the data symbols because the indent diagonal measurements were highly reproducible at each indentation load.

*S5. Indentation morphology*

Representative Vickers impressions for the smallest- and largest-grained specimens and a representative Berkovich nanoindentation array are shown in Figure S5. At 0.49 N, the Vickers impressions were well defined with limited cracking, whereas radial cracks extending from the indent corners were evident at 9.81 N. No obvious gross chipping or spallation was observed, and the overall Vickers indentation morphology was similar for S1750 and S1950. The Berkovich image qualitatively confirms the presence of well-spaced nanoindentations without obvious extensive surrounding surface damage. Because the vertical signal in this AFM-like image is reported in μN rather than calibrated height units, quantitative pile-up or sink-in could not be evaluated.

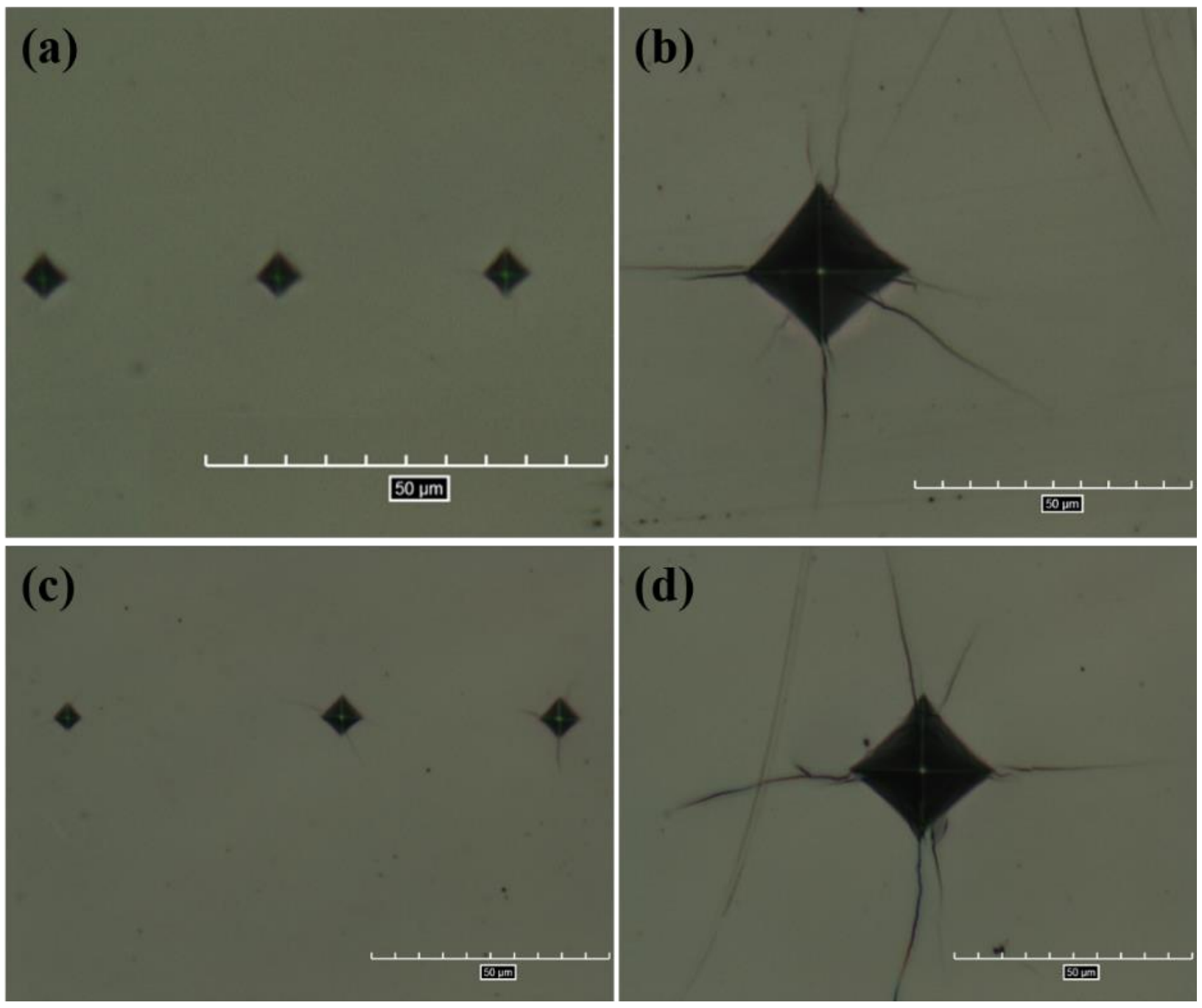

**(e)**

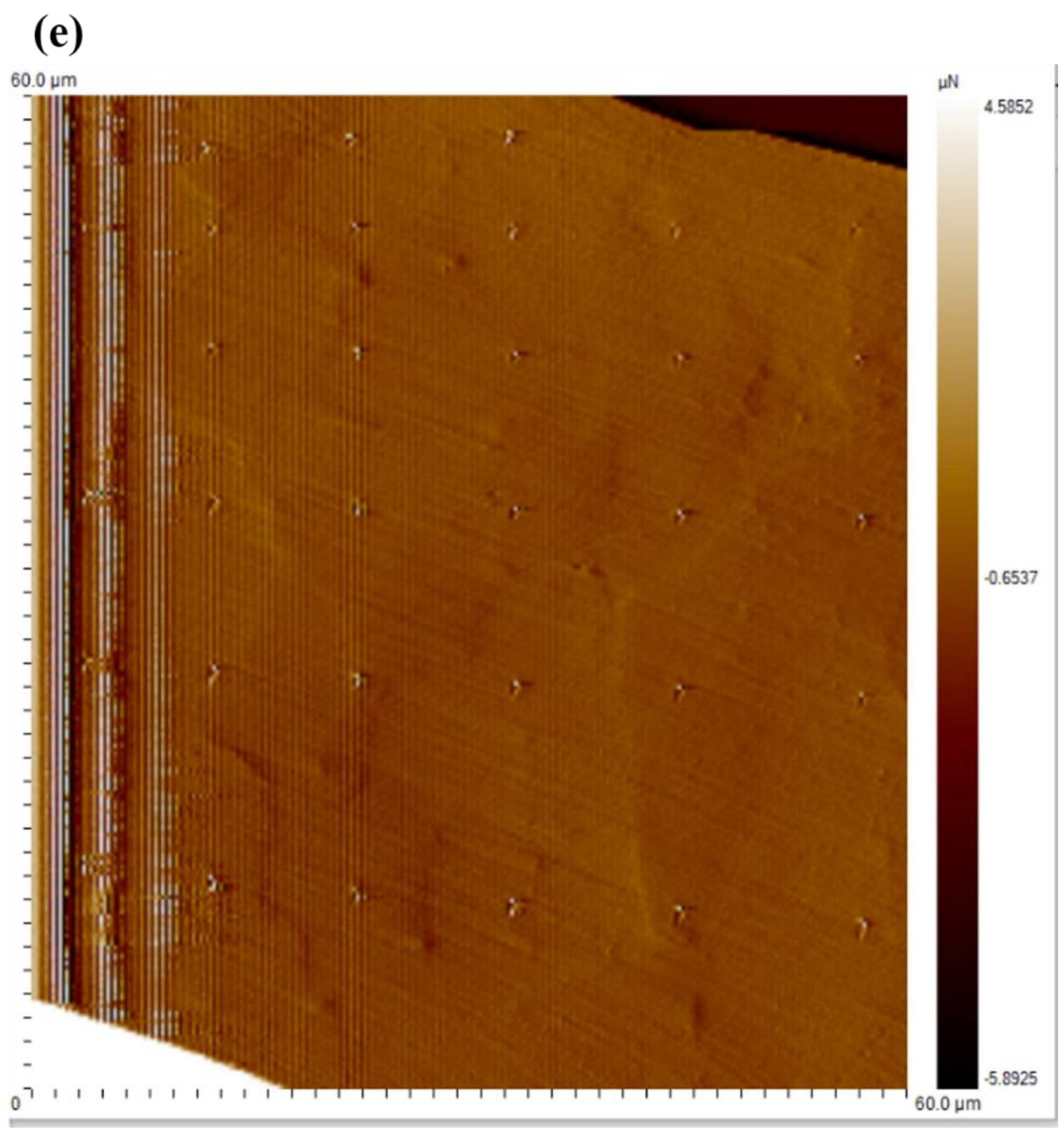

**Figure S5.** Representative indentation morphology of (Cr,Mo,Ta,V,W)$C_{1-\delta}$ ceramics: Vickers impressions for (a,b) S1750 and (c,d) S1950 at indentation loads of (a,c) 0.49 N and (b,d) 9.81 N, and (e) a representative Berkovich nanoindentation array obtained at 10 mN. The low-load Vickers impressions are well defined, whereas radial corner cracking is evident at 9.81 N, with no obvious gross chipping or spallation. The AFM-like nanoindentation image is shown for qualitative assessment of the indentation array and surrounding surface; its vertical signal is reported in µN and does not represent calibrated surface height.

*S6. Reduced modulus vs. grain size*

The reduced modulus measured by Berkovich nanoindentation varied from approximately 276 to 328 GPa across the grain-size series (Figure S6). The values showed the same general pattern as the nanohardness data, with lower values for the intermediate grain-size specimens and higher values for the smallest- and largest-grain-size specimens. However, the reduced modulus did not vary systematically with grain size, indicating that the elastic contact response was not controlled primarily by the threefold increase in grain size from S1750 to S1950.

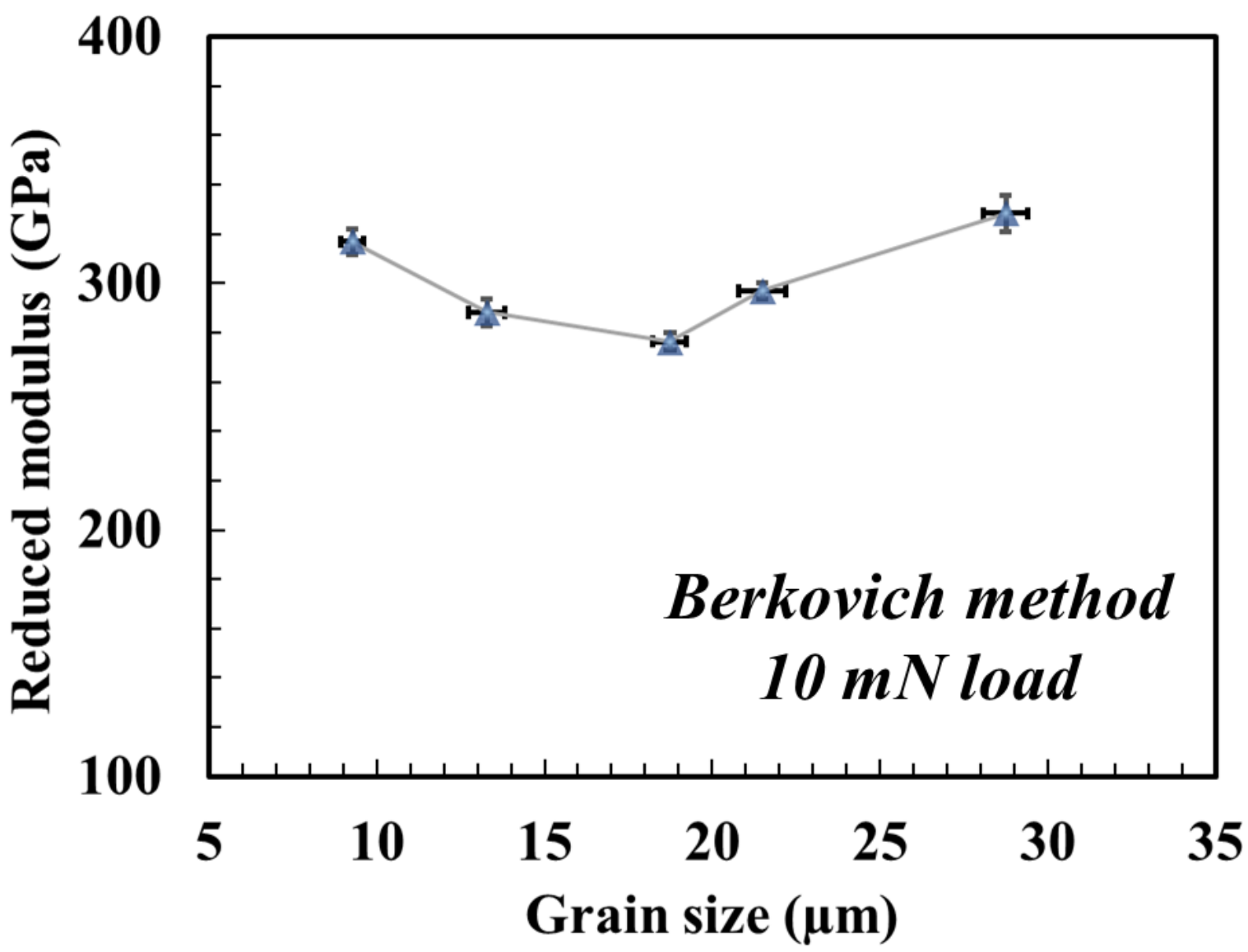

**Figure S6.** Reduced modulus as a function of grain size for fully dense (Cr,Mo,Ta,V,W)$C_{1-\delta}$ ceramics measured by Berkovich nanoindentation at a 10 mN load. The reduced modulus ranged from approximately 276 to 328 GPa and did not show a monotonic dependence on grain size.